\documentclass[epj]{svjour}
\usepackage{graphics}
\begin{document}
\title{$J/\psi$ production in PHENIX}
% \subtitle{Do you have a subtitle?\\ If so, write it here}
\author{Rapha\"el Granier de Cassagnac, for the PHENIX collaboration\inst{1} % etc
% \thanks{\emph{Present address:} Insert the address here if needed}%
}                     % Do not remove
\offprints{}          % Insert a name or remove this line
\institute{Laboratoire Leprince-Ringuet, \'Ecole
polytechnique/IN2P3, Palaiseau, 91128 France}
\date{Received: date / Revised version: date}
\abstract{ Heavy quarkonia production is expected to be sensitive to
the formation of a quark gluon plasma (QGP). The PHENIX experiment
has measured $J/\psi$ production at $\sqrt{s_{NN}}=$~200~GeV in
Au+Au and Cu+Cu collisions, as well as in reference  p+p and d+Au
runs. $J/\psi$'s were measured both at mid ($|y|<0.35$) and forward
($1.2<|y|<2.2$) rapidity. In this letter, we present the A+A
preliminary results and
 compare them to normal cold nuclear matter expectations derived from
PHENIX d+Au and p+p measurements as well as to theoretical models
including various effects (color screening, recombination,
sequential melting...).
\PACS{
      {PACS-key}{discribing text of that key}   \and
      {PACS-key}{discribing text of that key}
     } % end of PACS codes
} %end of abstract
\maketitle
PHENIX is one of the experiments located at the Relativistic Heavy
Ion Collider (RHIC) of Brookhaven National Laboratory. It has the
capability of measuring quarkonia through their dilepton decay in
four spectrometers: two central arms covering the mid-rapidity
region of $|\eta| < 0.35$ and twice $\pi/2$ in azimuth, and two
forward muon arms covering the full azimuth and $1.2 < |\eta| < 2.2$
in pseudorapidity. Electrons are identified in the central arms by
their {\v C}erenkov rings and by matching the momentum of charged
particles reconstructed in drift chambers with the energy deposited
in an electromagnetic calorimeter. In the forward arms, muons are
selected by an absorber and identified by the depth they reach in a
succession of proportional counters staggered with steel walls. The
event vertex and centrality are measured by beam-beam counters
situated at $3 < |\eta| < 3.9$. For A+A collisions, the centrality
measurement is further refined by the use of two zero degree
calorimeters located downstream the beams. A detailed description of
the PHENIX apparatus can be found in \cite{phenix:NIM}.

This letter is organized as follows: we first set up the references
by looking at d+Au data (section~\ref{sec:dA}) and compare the cold
nuclear matter effects derived from it to A+A data
(section~\ref{sec:Normal}). In section~\ref{sec:Anomalous}, we
compare our results to what was observed at the CERN SPS. We then
review in section~\ref{sec:Alternate} the state of the art of the
possible explanations of our observed suppression. Kinematic
distributions could help distinguishing between models and we
present our rapidity (section~\ref{sec:Rapidity}) and mean squared
transverse momentum (section~\ref{sec:MeanPt}) before to conclude in
section~\ref{sec:Final}.

Note that all A+A $J/\psi$ data shown below are PHENIX preliminary.

\section{$J/\psi$ production in d+Au collisions}
\label{sec:dA}

The nuclear modification factor $R_{AB}$ for any A+B collision type
is defined as the $J/\psi$ yield observed in these collisions,
divided by the yield measured in a p+p run and scaled by the average
number of nucleon-nucleon collisions $\langle N_{coll} \rangle$
extracted from a Glauber model:

\begin{equation}
R_{AB} = \frac{dN^{J/\psi}_{AB}}{\langle N_{coll}\rangle \times
dN^{J/\psi}_{pp}}
\end{equation}

\begin{figure}
\resizebox{0.45\textwidth}{!}{%
  \includegraphics{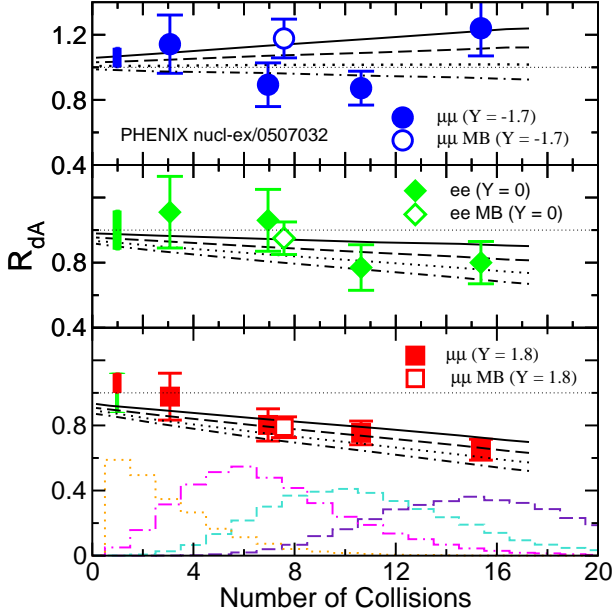}
} \caption{$J/\psi$ nuclear modification factor $R_{dA}$ as a
function of centrality (given here by the number of collisions
$N_{coll}$) for backward (top, $y=-1.7$) mid (middle, $y=0$) and
forward (bottom, $y=1.8$) rapidities. The histograms are the
distributions of $N_{coll}$ for each of our four centrality classes.
Theoretical curves are from Vogt~\cite{Vogt:dA} and take EKS
shadowing and 0, 1, 2 or 3~mb normal absorption cross section (from
top solid to bottom dot-dashed).}
\label{fig:dA}       % Give a unique label
\end{figure}

Departure from $R_{AB}=1$ implies some nuclear modifications of the
produced $J/\psi$'s. The PHENIX collaboration has measured the
$J/\psi$ nuclear modification factor for d+Au collisions
\cite{phenix:dA}. Its centrality dependence is shown on
figure~\ref{fig:dA} for our three rapidity ranges.

In d+Au collisions at 200 GeV, $J/\psi$'s in our three rapidity
ranges probe the following momentum fractions $x$ of gluons in the
gold nucleus (neglecting the emitted gluon): 0.05~to~0.14~(upper
panel, negative rapidity, gold-going side),
0.011~to~0.022~(midrapidity) and 0.0014~to~0.0047~(lower panel,
positive rapidity, deuteron-going side). The difference observed
between forward (lower panel) and backward (upper panel) yields
indicate that shadowing and/or anti-shadowing are at play. These
effects are modifications of the parton distribution functions in
nuclei with respect to the usual functions in free nucleons. In
particular, one expect a depletion (shadowing) at low $x$ due to
overlapping and recombining partons (gluons in the case of
$J/\psi$). The strength of gluon shadowing is not heavily
constrained by theory and models predictions
\cite{EKS,FGS,Kopeliovich} differ by a factor of three.

Comparison with Vogt's theoretical predictions~\cite{Vogt:dA} shows
that a modest amount of absorption (0 to 3~mb), added to a modest
amount of shadowing such as the one provided by the
Eskola-Kolhinen-Salgado (EKS~\cite{EKS}) scheme, can describe both
the rapidity and centrality dependencies, as shown on figure
\ref{fig:dA}.

\section{Normal nuclear matter effects in A+A}
\label{sec:Normal}

\begin{figure}
\resizebox{0.5\textwidth}{!}{%
  \includegraphics{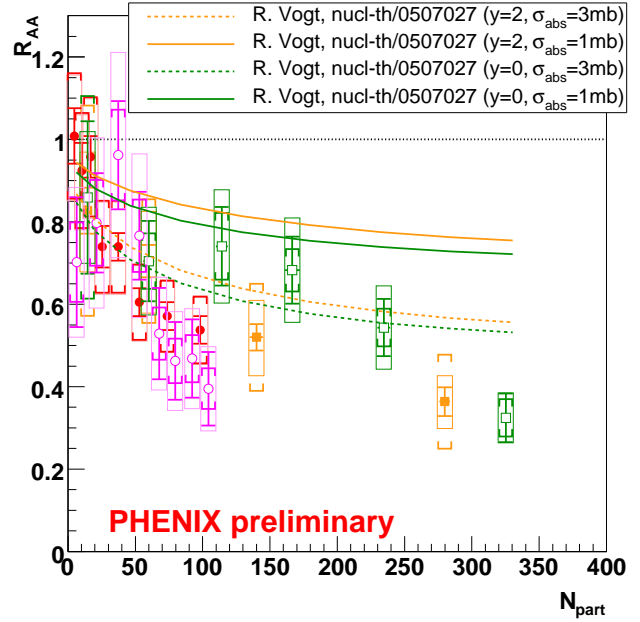}
} \caption{Nuclear modification factor $R_{AA}$ as a function of
centrality (given here by the number of participants $N_{part}$) for
Cu+Cu $y=0$ (open magenta circles) Cu+Cu $y=1.7$ (full red circles)
Au+Au $y=0$ (open green squares) and Au+Au $y=1.7$ (full gold
squares). Error bars are statistical, brackets are systematics, and
box are global (common to all the points). Theoretical curves are
Au+Au cold nuclear effects only predictions from
Vogt~\cite{Vogt:AA}, solid and dashed lines being for 1~mb and 3~mb
normal nuclear absorption cross sections, respectively.}
\label{fig:AAcold}       % Give a unique label
\end{figure}

To extrapolate the effects seen in d+Au to A+A collisions, one has
to rely on a model. Together with our Au+Au ($\sim$~0.2~nb$^{-1}$)
and Cu+Cu ($\sim$~3~nb$^{-1}$) measurements~\cite{phenix:AA}, Au+Au
predictions from such a cold nuclear matter model~\cite{Vogt:AA} are
presented on figure~\ref{fig:AAcold}, as a function of centrality
(displayed here as the number of participants). They account for
shadowing (following the EKS prescription) and nuclear absorption
(1~mb for solid lines and 3~mb for dashed lines). Predictions are
shown for two rapidity values ($y=0$ and $y=2$) corresponding to our
two rapidity regions. The difference between the 1~mb and the 3~mb
curves give an idea of our current uncertainty on the cold nuclear
matter effects affecting the $J/\psi$ production in A+A collisions.

However, in both rapidity ranges, our most central Au+Au
measurements depart from the predictions made with the stronger
nuclear effects (3~mb), suggesting that other suppression mechanisms
are involved, namely, an anomalous suppression.

\section{Lower energy anomalous suppression}
\label{sec:Anomalous}

\begin{figure}
\resizebox{0.5\textwidth}{!}{%
  \includegraphics{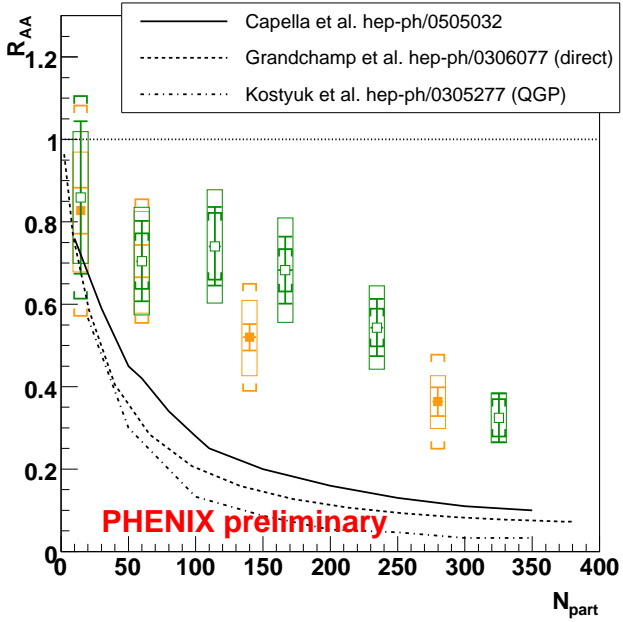}
} \caption{Nuclear modification factor $R_{AA}$ as already presented
on figure~\ref{fig:AAcold}. Only Au+Au data are shown ($y=0$ open
symbols, $y=1.7$ full symbols). Theoretical curves are predictions
from models that describe the anomalous suppression seen by the NA50
experiment at lower energies~\cite{Capella,Grandchamp,Kotsyuk}.}
\label{fig:AASPS}       % Give a unique label
\end{figure}

Such an anomalous suppression was early predicted by Matsui and Satz
as a signature of the QGP~\cite{Satz} and later observed by the CERN
NA50 experiment~\cite{NA50} at lower energy
($\sqrt{{s_{NN}}}=17.3$~GeV). Various models explain the NA50 data
and the extrapolation of three of them to RHIC energies are
displayed on figure~\ref{fig:AASPS} together with our Au+Au data.

In \cite{Capella} (solid lines), $J/\psi$'s are absorbed by comoving
particles (of undetermined partonic/hadronic nature). In
\cite{Grandchamp}, the authors describe the dynamical interplay
between suppression and regeneration of $J/\psi$'s in a QGP. The
suppression mechanism is dominant for NA50 energies and is the only
one presented here (see figure \ref{fig:AAReco} for the full
prediction). In \cite{Kotsyuk} (dot-dashed lines), the authors use a
QGP statistical charm coalescence model. All three models fail to
reproduce our data, overestimating the measured suppression. Other
models such as percolation~\cite{Percolation} also overpredicts the
suppression, suggesting that new mechanisms are at play at RHIC
energy.

\section{Alternate explanations for suppression }
\label{sec:Alternate}

For now, at least three classes of models exist that can roughly
accommodate the amount of anomalous suppression seen in our most
central preliminary results. The accuracy of the present data, as
well as the uncertainty on the cold nuclear effects, do not allow to
favor one or the other. We review them in the three following
subsections.

\subsection{Detailed transport}

\begin{figure}
\resizebox{0.5\textwidth}{!}{%
  \includegraphics{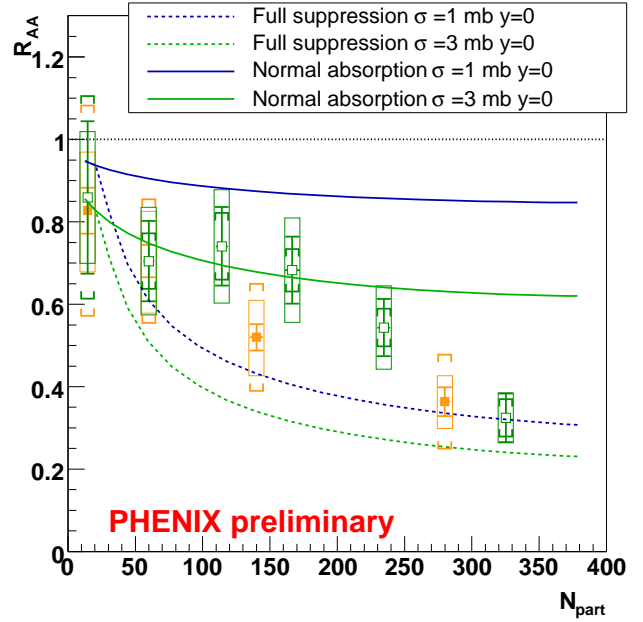}
} \caption{Nuclear modification factor $R_{AA}$ as already presented
on figure~\ref{fig:AAcold}. Only Au+Au data are shown ($y=0$ open
symbols, $y=1.7$ full symbols). Theoretical curves are from a
$J/\psi$ transport model in an hydrodynamical quark gluon
plasma~\cite{Zhu}. The dashed (solid) lines are with (without)
anomalous suppression. The up (down) curves are with 1~mb (3~mb)
normal nuclear absorption.}
\label{fig:AAZhu}       % Give a unique label
\end{figure}

One paper~\cite{Zhu}, simulating $J/\psi$ transport in a
hydrodynamical model, actually predicts an amount of suppression
that matches our most central data. It is shown on
figure~\ref{fig:AAZhu} where the authors have added cold nuclear
matter effects (nuclear absorption only, 1 or 3~mb) with respect to
the published paper. The suppression they obtain is not large
because they authorize $J/\psi$'s with sufficient momenta to freely
stream out of the plasma.

\subsection{Sequential melting}

 An important fraction (30 to 40~\%) of $J/\psi$'s comes
from feed-down decays of charmonia excited states ($\psi'$,
$\chi_c$). This particular point is taken care of in most of the
previous approaches. But recent lattice computations indicate that
$J/\psi$'s could melt at a much higher temperature than the one that
was originally thought. One possible hypothesis, defended
in~\cite{KKS}, is that only the excited states melt, leaving all the
initially produced $J/\psi$'s untouched. This 30 to 40~\%
suppression could also match our data.

\subsection{Recombination}

\begin{figure}
\resizebox{0.5\textwidth}{!}{%
  \includegraphics{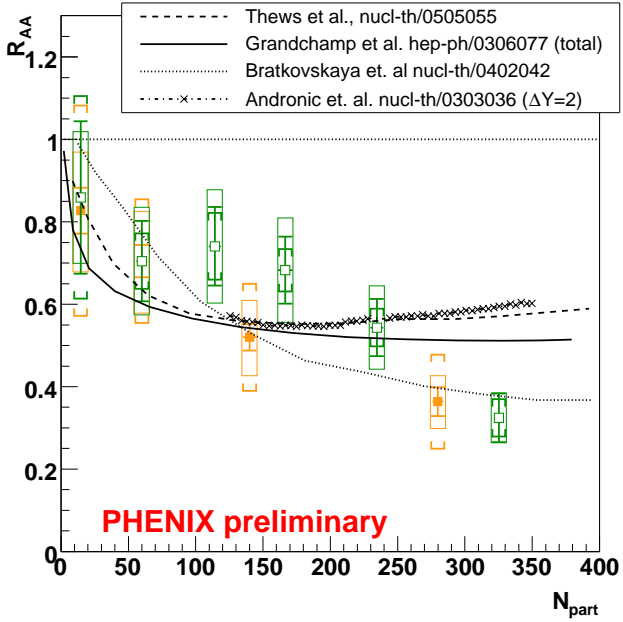}
} \caption{Nuclear modification factor $R_{AA}$ as already presented
on figure~\ref{fig:AAcold}. Only Au+Au data are shown ($y=0$ open
symbols, $y=1.7$ full symbols). Theoretical curves come from various
recombination models (solid \cite{Grandchamp}, dotted
\cite{Bratkoskaya}, crossed \cite{Andronic} and dashed
\cite{Thews}).}
\label{fig:AAReco}       % Give a unique label
\end{figure}

At RHIC energies, multiple $c\overline{c}$ pairs are produced (10 to
20 in central collisions according to~\cite{phenix:charm}). Quark
mobility in a deconfined medium could allow uncorrelated charm
quarks to recombine when the QGP fireball freezes, forming
quarkonia, and raising their yield with centrality. A balance
between suppression and enhancement could lead to the intermediate
suppression we observe. Figure~\ref{fig:AAReco} shows a collection
of predictions from various recombination or coalescence models,
from~\cite{Grandchamp,Bratkoskaya,Andronic,Thews}. Unfortunately,
the lack of knowledge of the yield and distributions of charm quarks
initially produced as well as of the recombination mechanism, make
these predictions hardly predictive. A good way to search for hints
of recombination is then to look at its impact on the distributions
of kinematical quantities, such as rapidity and transverse momenta.

\section{$J/\psi$ rapidity profile}
\label{sec:Rapidity}

\begin{figure}
\resizebox{0.5\textwidth}{!}{%
  \includegraphics{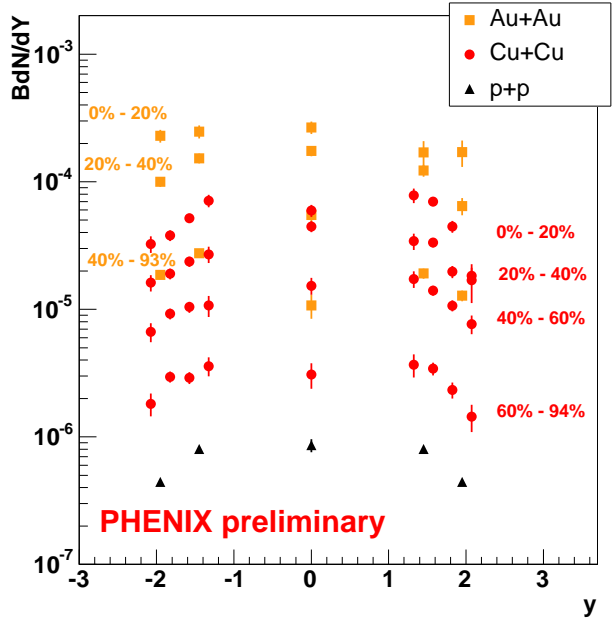}
} \caption{$J/\psi$ rapidity spectra for three Au+Au centrality
intervals (gold squares), four Cu+Cu centrality intervals (red
circles) and p+p collisions (black triangles).}
\label{fig:Rapidity}       % Give a unique label
\end{figure}

According to \cite{Thews}, the rapidity distribution of the
recombined $J/\psi$'s should be modified with respect to the
directly produced $J/\psi$'s. In particular, it should emphasize the
region of phase space where more $c$ quarks are produced, because
recombination probability goes quadratically with the $c$ quark
densities. Recombined $J/\psi$'s should then have a narrower
rapidity distribution. Figure~\ref{fig:Rapidity} shows our measured
rapidity spectra for three Au+Au and four Cu+Cu centrality classes,
as well as for p+p.

Within our limited statistical accuracy, we do not see any
modification of the rapidity profile and cannot draw any conclusion
towards or against recombination from this.

\section{$J/\psi$ mean transverse momentum}
\label{sec:MeanPt}

\begin{figure}
\resizebox{0.5\textwidth}{!}{%
  \includegraphics{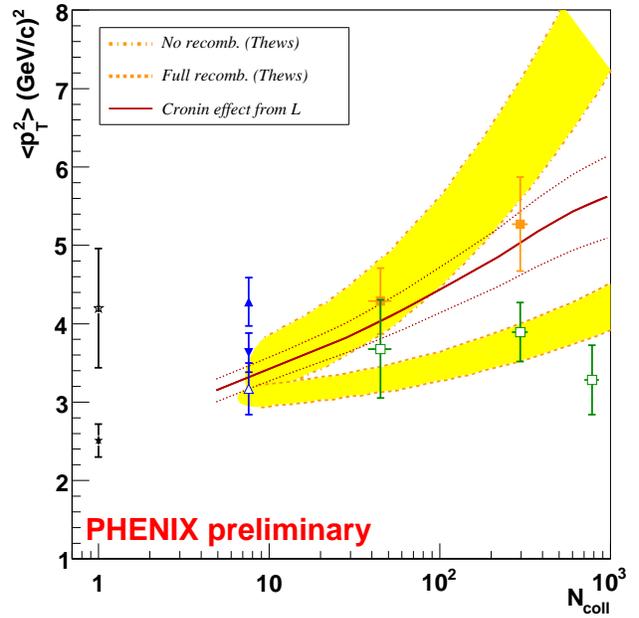}
} \caption{Mean squared transverse momentum versus $N_{coll}$. Black
stars at $N_{coll}=1$ are for p+p collisions, for mid (upper) and
forward (lower) rapidity. Triangles are for d+Au, for mid (open) and
forward/backward (full) rapidity. Squares are for Au+Au for mid
(open) and forward (full) rapidity. Yellow shaded bands are
from~\cite{Thews} and stands from direct (lower band) or fully
recombined $J/\psi$'s. The red line is a parametrization of Cronin
effect derived from d+Au data in~\cite{Tram}, the dotted lines on
each side reflecting the associated errors.}
\label{fig:pt2}       % Give a unique label
\end{figure}

Recombination modifies transverse momentum as well.
Figure~\ref{fig:pt2} shows our measured mean squared transverse
momentum $\langle p_T^2 \rangle$ for various systems and centrality
intervals, as a function of $N_{coll}$. The shaded yellow bands are
predictions from~\cite{Thews}, corresponding to either $J/\psi$'s
from recombination (lower band) or to directly produced $J/\psi$'s
(upper band). To properly predict the modified $p_T$ spectra, one
first needs to quantify the normal $p_T$ broadening coming from cold
nuclear effects (Cronin effect). This effect was clearly seen by
PHENIX~\cite{phenix:dA} at forward rapidity, by comparing $\langle
p_T^2 \rangle$ in p+p and d+Au collisions. In~\cite{Thews} the
authors use these measurements to quantify the Cronin effect.
However, at midrapidity, because of the very poor statistics of the
p+p sample, it is difficult to quantify it. Thus, the theoretical
prediction can only be safely compared to the forward rapidity case.

The solid red line is another parametrization of the Cronin effect
from~\cite{Tram}, using this simple parametrization:
\begin{equation}
\langle p_T^2 \rangle_{AA} = \langle p_T^2 \rangle_{pp} + \rho
\sigma \delta(\langle p_T^2 \rangle) \times L = 2.51 + 0.32 \times L
\end{equation}
where $\langle p_T^2 \rangle$ is in units of $(GeV/c)^2$ and $L$ is,
in fermis, the average thickness of nuclear matter seen by a
$J/\psi$ in the collision system and the centrality bin considered.
The factor $\rho \sigma \delta(\langle p_T^2 \rangle) = 0.32$ is
determined by the d+Au forward data and stands for the nuclear
density $\rho$, times the elastic gluon-nucleon scattering cross
section $\sigma$, times the average $p_T$ kick given at each
scattering $\delta(\langle p_T^2 \rangle)$. The dotted lines on each
side of the main curve are the errors derived from the p+p and d+Au
uncertainties. This parametrization provided a good description of
lower energy experiments~\cite{Tram} and matches our data when it
can be safely applied, namely at forward rapidity.

Thus, the raise of $\langle p_T^2 \rangle$ at forward rapidity in
Au+Au collisions seems explainable by the Cronin effect only. A
better $p+p$ baseline is needed to interpret the rather centrality
independent $\langle p_T^2 \rangle$ we observe at midrapidity .

\section{Conclusions}
\label{sec:Final}

PHENIX preliminary results show a substantial amount of $J/\psi$
suppression, growing with centrality. At the highest energy density
probed so far, this suppression is as large as a factor of three.
However, its interpretation is not easy. First, the amount of
normal, cold nuclear matter suppression is not precisely known and
demands more d+A data. Nevertheless, the observed suppression in A+A
seems to exceed the maximum suppression authorized by cold nuclear
matter, and thus be partly anomalous.

It also seems weaker than the suppression derived from models that
fit lower energy data. Three classes of models that raise the number
of surviving $J/\psi$'s were presented. They all suppose the
formation of a QGP. To distinguish between them, a better precision
on data, and in particular on the kinematical distributions, is
required.

% \subsection{J'ai droit au subsections}
% \label{sec:2} and \cite{RefJ} as required. Don't forget to give each
% section and subsection a unique label (see Sect.~\ref{sec:1}).
%\begin{table}
%\begin{tabular}{lll}
%\hline\noalign{\smallskip}
%\end{tabular}
%% Or use
%\vspace*{5cm}  % with the correct table height
%\end{table}
%
% BibTeX users please use
% \bibliographystyle{}
% \bibliography{}

\end{document}